# High-pressure Raman spectroscopy and lattice-dynamics calculations on scintillating MgWO$_4$: A comparison with isomorphic compounds

J. Ruiz-Fuertes[1,†], D. Errandonea[1], S. López-Moreno[2], J. González[3,4], O. Gomis[5], R. Vilaplana[5], F.J. Manjón[6], A. Muñoz[2], P. Rodríguez-Hernández[2], A. Friedrich[7], I. A. Tupitsyna[8], and L. L. Nagornaya[8]

[1] Departamento de Física Aplicada-ICMUV, MALTA Consolider Team, Universitat de València, Edificio de Investigación, c/Dr. Moliner 50, 46100 Burjassot, Spain.

[2] Departamento de Física Fundamental II, MALTA Consolider Team, Departamento de Física Fundamental II, Instituto de Materiales y Nanotecnología, Universidad de La Laguna, La Laguna, 38205 Tenerife, Spain.

[3] DCITIMAC, MALTA Consolider Team, Universidad de Cantabria, Avda. de Los Castros s/n, 39005 Santander, Spain.

[4] Centro de Estudios de Semiconductores, Universidad de los Andes, Mérida 5201, Venezuela.

[5] Centro de Tecnologías Físicas: Acústica, Materiales y Astrofísica, MALTA Consolider Team, Universidad Politècnica de València, Camino de Vera s/n, 46022 Valencia, Spain.

[6] Instituto de Diseño para la Fabricación y Producción Automatizada, MALTA Consolider Team, Universidad Politécnica de Valencia, Camino de Vera s/n, 46022 Valencia, Spain.

[7] Institut für Geowissenschaften, Abt. Kristallographie, Goethe-Universität Frankfurt, D-60438 Frankfurt am Main, Germany.

[8] Institute Scintillating Materials, UA-61001 Kharkov, Ukraine.

**Abstract:** Raman scattering measurements and lattice-dynamics calculations have been performed on magnesium tungstate under high pressure up to 41 GPa. Experiments have been carried out under a selection of different pressure-media. The influence of non-hydrostaticity on the structural properties of MgWO$_4$ and isomorphic compounds is examined. Under quasi-hydrostatic conditions a phase transition has been found at 26 GPa in MgWO$_4$. The high-pressure phase has been tentatively assigned to a triclinic structure similar to that of CuWO$_4$. We also report and discuss the Raman symmetries, frequencies, and pressure coefficients in the low- and high-pressure phases. In addition, the Raman frequencies for different wolframites are compared and the variation of the mode frequency with the reduced mass across the family is investigated. Finally, the accuracy of theoretical calculations is systematically discussed for MgWO$_4$, MnWO$_4$, FeWO$_4$, CoWO$_4$, NiWO$_4$, ZnWO$_4$, and CdWO$_4$.

---

[†] e-mail address: javier.ruiz-fuertes@uv.es

# I. INTRODUCTION

Divalent-metal tungstates ($AWO_4$) are currently being studied with great interest due to their use as materials for scintillator detectors, laser-host crystals, as well as in acoustic and optical fiber applications [1, 2]. Magnesium tungstate ($MgWO_4$), the mineral hunzalaite, is part of this family. With a band gap of 3.92 eV [3], it is one of the most extensively studied metal tungstates because of its interest as a scintillator material for cryogenic applications, used in search for rare events in particle physics [4]. $MgWO_4$ crystallizes in a monoclinic structure isomorphic to wolframite, like other tungstates as $MnWO_4$, $FeWO_4$, $CoWO_4$, $NiWO_4$, $ZnWO_4$, and $CdWO_4$. It belongs to space group $P2/c$ (SG: 12) and has $C_{2h}$ point-group symmetry. The structure consists of layers of alternating $AO_6$ (A = Mg, Mn, Fe, Co, Ni, Zn, Cd) and $WO_6$ octahedral units that share edges forming a zig-zag chain [5] and creating a close-packed structure. The crystalline structure is illustrated in Fig. 1.

Several studies of the physical properties of $MgWO_4$ have been carried out at ambient pressure. In particular, the electronic, vibrational, and scintillating properties have been studied [3, 4, 6]. High-pressure research has proven to be an efficient tool to improve the understanding of the main physical properties of compounds related to $MgWO_4$. However, only a limited number of high-pressure (HP) studies have been performed on this compound. After the single-crystal x-ray diffraction work of Macavei and Schulz [7] where three wolframites were studied up to 9 GPa, a few works have appeared on the high-pressure vibrational [8, 9, 10] and structural [6, 10, 11] properties of $MgWO_4$ and other wolframites. The high-pressure behavior of $MgWO_4$ was expected to be very similar to that of $ZnWO_4$ which shows only one structural phase transition at 30.6 GPa according to Raman experiment and *ab initio* calculations [8]. However XRD experiments carried out on both $MgWO_4$ and $ZnWO_4$ showed that these wolframites undergo, in addition to the expected phase transition around 31 GPa, another one around 17 GPa [6]. From combined Raman spectroscopic and *ab initio* calculations [8, 9] it has been established that the monoclinic β-fergusonite structure (space group *C*2/*c*, SG: 13) could be the most probable HP phase for these compounds. In addition, a triclinic structure (space group $P\bar{1}$, SG: 2) similar to that of CuWO$_4$ is energetically competitive with the wolframite and β-fergusonite phases. This fact has been used to explain the additional phase transition observed by means of powder XRD in MgWO$_4$ and ZnWO$_4$.

In order to further understand and explain the structural behavior of wolframites under compression we have performed a combined Raman spectroscopy and theoretical study of MgWO$_4$ up to 41 GPa. Experiments were carried out using various pressure-transmitting media (PTM) including neon, methanol-ethanol, spectroscopic paraffin, and no PTM at all. This study will provide us not only with a better knowledge of its vibrational properties but also the possibility to correlate the general trends of wolframites at high pressure.

## II. EXPERIMENTAL DETAILS

Five series of Raman measurements were performed on 10 μm-thick platelets cleaved from MgWO$_4$ single crystals (or with micron-sized powders grounded from the crystals) using three different Raman spectrometers in backscattering geometry. Single crystals were prepared using the flux growth technique developed at the Institute for Scintillation Materials (Kharkov, Ukraine) [12]. A stoichiometric mixture of MgO and WO$_3$ (99.99%) was added to the flux prepared from Na$_2$WO$_4$ (99.95%) at 790 °C in a platinum crucible. Single-crystalline samples of MgWO$_4$ crystals of circa 1 cm$^3$ were grown by pulling the seed from the melted flux solution.

The first experiment was carried out using a Renishaw Raman spectrometer (RM-1000) with a 1800 grooves/mm grating, 100 μm slit and equipped with a HeNe laser (633 nm, 50 mW), a 20x objective (1 cm$^{-1}$ of spectral resolution) and a nitrogen ($N_2$) cooled charge-coupled device (CCD) detector. The sample was loaded in a Boehler-Almax diamond-anvil cell (DAC) with neon as pressure medium and a maximum pressure of 40 GPa was reached. The second and third experiments were carried out using a LabRam HR UV microRaman spectrometer with a 1200 grooves/mm grating, 100 μm slit and a 50x objective (spectral resolution of 2 cm$^{-1}$), in combination with a thermoelectric-cooled multichannel CCD detector. A 532.12 nm laser line with a power of 10 mW was used. In these experiments we went up to 41 (31) GPa and used a Boehler-Almax (membrane-type) DAC loaded with neon (no PTM). In the experiment performed without PTM we used $MgWO_4$ powder. The fourth and fifth experiments were carried out using a triple monochromator Jobin-Yvon T64000 in the subtractive mode with a resolution of 0.8 cm$^{-1}$, a 1800 grooves/mm grating, 100 μm slit and equipped with a liquid $N_2$ cooled CCD detector with a 514.5 nm line of an argon laser focused down with a 20x objective and keeping the power on the sample below 5mW, in order to avoid laser-heating effects on the probed material and the concomitant softening of the observed Raman peaks. In the fourth experiment pressure was increased up to 21 GPa in a membrane-type DAC using a 4:1 mixture of methanol-ethanol as PTM and in the fifth one up to 30 GPa using $MgWO_4$ powder with purity higher than 99.5 % (Mateck), a membrane DAC and spectroscopic paraffin as PTM. In all experiments pressure was determined suing the ruby-fluorescence technique [13] (with ±1% maximum uncertainty). In addition we collected Raman spectra of $NiWO_4$, $CoWO_4$, and $FeWO_4$ (the mineral ferberite) at ambient pressure in order to compare them with theoretical calculations. Powders of 99.9 % purity (Alfa Aesar) were used in the experiments.

**III. CALCULATIONS DETAILS**

In the last years *ab initio* methods have allowed detailed studies of the energetics of materials under high pressures [14]. In this work total-energy calculations were done within the framework of the density-functional theory (DFT), the Kohn-Sham equations were solved using the projector-augmented wave (PAW) [15,16] method as implemented in the Vienna *ab initio* simulation package (VASP) [17]. We used a plane-wave energy cutoff of 520 eV to ensure accurate and high precision in the calculations. The exchange and correlation energy was described within the GGA in the PBE [18] prescription for $MgWO_4$. The Monkhorst-Pack (MP) [19] grid used for Brillouin-zone integrations ensured highly converged results for the analyzed structures (to about 1 meV per formula unit). It has been pointed out in different studies of transition metal compounds that GGA often yields incorrect results for systems with high correlated electrons. The implementation of the DFT+$U$ method has been found to have some influence on transition metal compounds [20]. The GGA+$U$ method was used to account the strong correlation between the electrons in the *d* orbitals on the basis of Dudarev's method [20] for the study of $AWO_4$ (A= Mn, Co, Ni and Fe). In this method the on-site Coulomb interaction, $U$ (Hubbard term), and the on-site exchange interaction, $J_H$, are treated together as $U_{eff} = U - J_H$. For our GGA+$U$ calculations, we have chosen a value $U_{eff}$ = 3.9, 4.2, 4.3, and 7 eV for Mn, Co, Ni, and Fe atoms, respectively. For these compounds we have performed spin density calculations and we found that the antiferromagnetic configuration was the stable one for Mn, Co, Ni, and Fe wolframites. In the relaxed equilibrium configuration, the forces are less than 6 meV/Å per atom in each of the cartesian directions. Lattice-dynamics calculations of phonon modes were performed at the zone centre (Γ point) of the BZ. The calculations provided information about the frequency, symmetry and polarization vector of the vibrational modes in each structure. Highly converged results on forces are required for the calculation of the dynamical matrix from lattice-dynamics calculations. We used the direct force-constant approach (or supercell method) [21]. The construction of the dynamical matrix at the Γ point of the BZ is particularly simple and involves separate calculation of the forces in which a fixed displacement from the equilibrium configuration of the atoms within the primitive unit cell is considered. Symmetry further reduces the computational efforts by reducing the number of such independent displacements in the analyzed structures. Diagonalization of the dynamical matrix provides both the frequencies of the normal modes and their polarization vectors. It allows us to identify the irreducible representations and the character of phonon modes at the Γ point.

**IV. RESULTS AND DISCUSSION**

**A. Low pressure phase**

According to group-theory analysis the wolframite structure has 36 vibrational modes, 18 being Raman active (even vibrations *g*) and 18 infrared (IR) active (odd vibrations *u*) at the Γ point: $\Gamma = 8A_g + 10B_g + 8A_u + 10B_u$. The assignment of the modes is shown in Table I [6]. It has been argued that for wolframite-type *A*WO$_4$ compounds the Raman modes can be classified as internal and external modes with respect to the WO$_6$ octahedra [8, 9]. Thus wolframites have up to six internal stretching modes that arise from the six W-O bonds in the WO$_6$ octahedra. Since W is heavier than Mg and W-O covalent bonds are stiffer than Mg-O bonds, it is reasonable to think of the material as two separated blocks, one concerning the WO$_6$ units and the second one the Mg$^{2+}$ cation. Moreover, it is known that WO$_6$ octahedra are quite incompressible, the MgO$_6$ octahedra accounting for most of the volume reduction of the structure under pressure. Consequently, the six internal stretching modes of MgWO$_4$ are in the high-frequency part of the Raman spectrum. They are the A$_g$ modes with frequencies 917, 713, 552, and 420 cm$^{-1}$ and the B$_g$ modes with frequencies 809 and 684 cm$^{-1}$.

Figures 2, 3a, and 3b show the Raman spectra of MgWO$_4$ at some selected pressures up to 41.0, 30.6 and 30.2 GPa, respectively. Data were collected using different PTM like neon, no PTM, and spectroscopic paraffin, respectively. Depending upon the experiment, we can follow all the Raman modes of wolframite up to 17 - 29 GPa when the clear appearance of an additional mode just below the most energetic one, plus the intensity drop of several of the wolframite modes indicate the onset of a phase transition. Figure 4 shows the evolution of the measured Raman modes as obtained from the experiments with Ne and methanol-ethanol, which are in good agreement. Since the Raman modes evolve linearly with pressure we have obtained the pressure coefficients (d$\omega$/d$P$) of them by means of linear fits. HP and LP [6] results are summarized in Table I together with results from *ab initio* calculations. The mode assignment stated in Table I is supported by theory and polarized Raman measurements.

It is interesting to point out how well both the experimental and the calculated modes agree, being the major difference for most of them less than 10 cm$^{-1}$. This good agreement also accounts for their pressure coefficients except for that of the B$_g$ mode at 405 cm$^{-1}$. For this mode the experimental pressure coefficient is 5 times lower than the calculated one. From the pressure coefficients of the mode frequencies and the bulk modulus $B_0$ we have obtained the Grüneisen parameters, $\gamma = (B_0/\omega) \cdot (d\omega/dP)$, for MgWO$_4$ (Table I). For the experimental Grüneisen parameters the value for $B_0$ (160 GPa) was taken from Ref. 6, while for the calculated data the theoretical value of $B_0$ (161 GPa) was used for self-consistency. Both the experimental and calculated Grüneisen parameters match very well with each other for the low-pressure phase with only a mismatch shown again for the 405 cm$^{-1}$ mode. This discrepancy has not been previously observed for other wolframites ($ZnWO_4$ and $CdWO_4$) and its origin remains unclear.

In all the experiments the behavior of the modes upon compression is quite similar up to 10 GPa. However, differences start to appear at higher pressures in the experiments performed without PTM or with paraffin. If no PTM is used, most modes show a faster frequency increase beyond 10 GPa than in experiments using Ne. One example is the $A_g$ mode with frequency 294 cm$^{-1}$. Only for few modes, like the 420 cm$^{-1}$ $A_g$ mode, the pressure behavior is the same in all experiments. If paraffin is used as the PTM, all modes have a larger pressure coefficient than in the rest of the experiments beyond 10 GPa. The origin of these differences can be caused by the presence of non-negligible uniaxial stress in the experiments performed without PTM or with paraffin [22]. Consequently, these two experiments were not used to obtain the results summarized in Fig. 4 and Table I.

By means of the harmonic approximation, if we consider that the atoms are bonded by means of springs then it can be stated that the frequency of the oscillations is directly proportional to the inverse square root of the reduced mass of the cations. In our case for simplicity we will consider that our system consists on two separate blocks one being the cation A and the other being the anion $WO_4$. Thus, in order to identify some general trends on $AWO_4$ wolframites, we have plotted the Raman shifts of the different vibrational modes of the wolframite series as a function of the inverse of the square root of the reduced mass, µ, of the A cation and the $WO_6$ polianion ($1/µ = 1/m_A + 1/m_{WO6}$) in Figs. 5 and 6. Table II summarizes the experimental [6, 8, 9, 10, 23, 24] and calculated Raman modes and pressure coefficients (in parenthesis) for the whole wolframite family. The first conclusion we obtained is that the seven compounds have a similar overall mode distribution. As expected, internal high-frequency modes, in which O atoms vibrate against W atoms, are very close for all five compounds and show little dependence of the mass of the $A^{2+}$ cation. On the contrary, the external low-frequency modes, which involve motions of $WO_6$ polyhedra against the A cation, are more sensitive to the mass of the divalent cation. In particular, we found that $MgWO_4$, $ZnWO_4$, and $CdWO_4$ follow a systematic trend and $MnWO_4$, $FeWO_4$, $CoWO_4$, and $NiWO_4$ another one. As can be seen in Table II and Figs. 5 and 6, the wolframites which do not involve magnetic cations ($MgWO_4$, $ZnWO_4$, and $CdWO_4$) show an inverse proportional relation between the frequencies of the external modes and the square root of the reduced mass $\mu$. In particular, the $B_g$ mode located at 405 cm$^{-1}$ for $MgWO_4$ is extremely sensitive to the mass of the divalent cation. Indeed, in Fig. 5 and Table II it can be seen that the mode-frequency ($\nu$) sequence in the 350 – 400 cm$^{-1}$ region changes from $\nu_{Bg} > \nu'_{Bg} > \nu_{Ag}$ in $MgWO_4$ to $\nu'_{Bg} > \nu_{Ag} > \nu_{Bg}$ in $CdWO_4$ (the quotation mark is used to differentiate between different $B_g$ modes). We would like to note here that these three modes have very similar pressure coefficients in the three compounds. It is also interesting that the influence of the atomic mass of the divalent cation on the phonon frequencies of the external modes, which we observed in non-magnetic wolframites, is similar to that found in alkaline-earth tungstates [25]. In contrast, the external-mode frequencies of $MnWO_4$, $FeWO_4$, $CoWO_4$, and $NiWO_4$ show the opposite behavior as they increase with the divalent-cation mass. This different behavior could be caused by the influence of magnetic interactions and second-order Jahn-Teller effects which induce strong distortions of the $WO_6$ and $AO_6$ octahedra. Actually these effects could even become strong enough to induce triclinic distortion, as is the case for CuWO$_4$ [10]. It is interesting to note that the Raman spectrum of wolframite-type CuWO$_4$, which is obtained at 10 GPa after undergoing a phase transition, resembles that of magnetic wolframites (see Table II) [10]. This fact supports the hypothesis described above. Nevertheless, the discussion of the influence of magnetic and Jahn-Teller effects on the lattice vibrations of wolframites is beyond the scope of this paper.

In the case of MgWO$_4$, ZnWO$_4$, and CdWO$_4$ a few differences are observed that we would like to highlight. For example, there is a frequency gap between the less energetic mode (97 cm$^{-1}$) and the following one (156 cm$^{-1}$), that happens to be higher for MgWO$_4$ than for the other members of the wolframite family. Further, the A$_g$ (277 cm$^{-1}$) mode moves more slowly with pressure than the B$_g$ (294 cm$^{-1}$) mode for MgWO$_4$, while the opposite happens for ZnWO$_4$ and CdWO$_4$. Another difference is related to the pressure coefficient of the B$_g$ mode located near 385 cm$^{-1}$ for MgWO$_4$ and at 354 and 352 cm$^{-1}$ for ZnWO$_4$ and CdWO$_4$, respectively, which is around 4 times higher than that of the surrounding modes. Hence, this mode crosses other modes at high pressures as it can be seen in Fig. 4 as well as in Refs. 8 and 9. Finally, the pressure coefficients of the four internal modes at higher frequencies are slightly lower for MgWO$_4$ than for ZnWO$_4$ and CdWO$_4$. This observation is consistent with the fact that MgWO$_4$ is the least compressible compound among the wolframite family.

**B. *Ab initio* calculations**

*Ab initio* calculations usually describe well the HP structural properties of wolframites [6, 8, 9]. In particular, calculations have been performed for MgWO$_4$, CdWO$_4$, ZnWO$_4$, and MnWO$_4$. Here we report structural calculations for FeWO$_4$, NiWO$_4$, and CoWO$_4$. For the three compounds, a wolframite-type structure is found to be the stable structure at ambient pressures. Since these compounds could be magnetic due to the presence of the $Co^{2+}$, $Fe^{2+}$, and $Ni^{2+}$ cations, we considered different magnetic configurations. We found the low-pressure phase of the three compounds to have a wolframite structure with an antiferromagnetic configuration. In this configuration, Mn, Fe, Co, and Ni have a magnetic moment of 4.3 μB, 3.8 μB, 2.7 μB, and 1.65 μB, respectively. These magnetic structures agree with neutron-diffraction studies and x-ray absorption experiments [26 – 28]. The obtained magnetic order and moments are also comparable with results reported for antiferromagnetic $MnWO_4$ [23]. In Table III we summarize the calculated structural parameters and compare them with experimental values [29, 30]. The agreement for the lattice parameters and the atomic coordinates is good with an underestimation of the unit-cell volume for $CoWO_4$ and $NiWO_4$ of 4.8 and 4.2% and an overestimation for $FeWO_4$ of 3.5%. These differences are typical for density-functional theory calculations [31, 32].

In addition, to the Raman-active phonons of $MgWO_4$, we have also calculated them for $MnWO_4$, $FeWO_4$, $CoWO_4$, and $NiWO_4$. In Table II, it can be seen that at ambient pressure the agreement of calculations with experiment is good. Raman-mode assignment has been done based upon calculations. The internal modes have been identified and are depicted by asterisks in the table. For $NiWO_4$, the present DFT calculations agree much better with experiments than previous calculations performed using the periodic linear-combination of atomic orbitals method [33]. In particular, calculations gave excellent agreement for the internal modes; discrepancies are always smaller than 5%. As was discussed above, for $MgWO_4$ the agreement is not only good for ambient pressure frequencies, but also for pressure coefficients. The same can be stated for $ZnWO_4$ and $CdWO_4$ [8, 9]. Therefore, *ab initio* calculations show to be an efficient tool to characterize the lattice dynamics of wolframites at ambient and high-pressure. Only for the 405 $cm^{-1}$ external $B_g$ mode the discrepancies in the pressure coefficient are important. This mode is the same one that is extremely sensitive to the mass of the divalent cation, as discussed in the previous section. For NiWO$_4$, CoWO$_4$, and FeWO$_4$, there have no HP Raman measurements been performed yet. Consequently, given the good description provided for other wolframites by *ab initio* calculations, we calculated the pressure evolution of Raman phonons for MnWO$_4$, NiWO$_4$, CoWO$_4$, and FeWO$_4$. Results are shown in Table II, where the pressure coefficients are included. Like for other wolframites the pressure coefficient is larger for the internal modes than for the rest of the modes. Within the internal modes the B$_g$ modes are those more sensitive to pressure. Also, the external modes with the highest frequencies (two B$_g$ modes) are very sensitive to pressure. On the other hand, there is a low-frequency A$_g$ mode with frequency between 124 cm$^{-1}$ and 141 cm$^{-1}$ which in the three compounds has an extremely small pressure coefficient. The triclinic wolframite-type CuWO$_4$ shows a similar pressure evolution of the phonon frequencies as MnWO$_4$, NiWO$_4$, CoWO$_4$, and FeWO$_4$ (Table II). To conclude this section, it is interesting to note that in contrast with scheelite-structured oxides [25] no phonon softening occurs in wolframites upon compression.

**C. High-pressure phase**

As was already mentioned in section A, the occurrence of a phase transition is observed by the appearance of an additional Raman mode at a wavelength slightly smaller than the most intense mode of wolframite at different pressures between 17 and 30 GPa depending on the PTM used (Figs. 2, 3). The phase transition is reversible with little hysteresis in all the experiments. In the experiment performed using Ne as PTM (Fig. 2), the appearance of the new mode at 25.8 GPa is followed by a quick increase of its intensity and the appearance of extra Raman bands. A total of 18 emerging modes are observed at 38 GPa. Simultaneously, a decrease in relative intensity of the other modes is observed, which fully disappear at 38 GPa. In the experiment performed without using any pressure medium (Fig. 3 (a)), the same process happens at a higher pressure of 29.6 GPa but more gradually. Finally, in the experiment with spectroscopic paraffin (see Fig. 3 (b)), the HP modes become evident as early as at 17 GPa and the low-pressure modes remain still observable up to 30 GPa, the maximum pressure of the experiment. Phase coexistence is found in all three experiments and the pressure range of coexistence depends upon the PTM used. In order to quantify the gradual transformation, we have analyzed the effect of pressure on the intensities of distinctive modes of both phases and calculated the intensity ratio $I_{HP}/(I_{LP}+I_{HP})$ [34]. In this equation, $I_{HP}$ is the intensity of the highest-frequency mode of the HP phase whereas $I_{LP}$ is the highest-frequency mode of wolframite. Both modes are the strongest of each structure and they do not overlap in the pressure range of coexistence, allowing a reasonable estimation of the HP/LP phase proportion. These modes are depicted by two arrows in Figs. 2, 3(a), and 3(b). The results are plotted in Fig. 7, there it can be seen that in experiments using Ne a steep increase of the intensity ratio from 0 at 26 GPa (only low-pressure phase) to 1 at 36 GPa (only high-pressure phase). In contrast, in the experiment using paraffin the changes are detected at 17 GPa, reaching the intensity ratio 0.8 at 30 GPa. This indicates that at 30 GPa there are still domains of the low-pressure phase present. From Fig. 5, it can be extrapolated that the intensity ratio would reach 1 at around 36 GPa as is the case where Ne was the PTM. These results in combination with the appearance of the same bands after the phase transition in all the experiments indicate that the mechanism of the phase transition is the same for all of them. However, the onset pressure depends on the hydrostaticity of the PTM. The fact that the transition is detected at the lowest pressure when the experiment is performed with paraffin suggests that beyond 10 GPa paraffin becomes stiffer than the wolframites. A similar behavior has already been observed previously for ZnWO$_4$ [6, 8] indicating that non-hydrostatic conditions in wolframites accelerate the transition onset.

The effect of non-hydrostaticity on MgWO$_4$ becomes further visible from the analysis of the pressure dependence of the full-width at half maximum (FWHM) for some Raman modes. In Fig. 8 we show the results for two modes of MgWO$_4$, the A$_g$ mode at 916.8 cm$^{-1}$ (a) and the B$_g$ mode at 97.4 cm$^{-1}$ (b), under four different experimental conditions. It can be concluded that wolframites pressurized without any PTM suffer uniaxial stresses that are important even at low pressures as evidenced by the strong peak broadening. In the experiment using paraffin we found a steep increase of the FWHM for both modes beyond 5 GPa. At this pressure, bands are as broad as without PTM indicating that experimental conditions are far away from quasi-hydrostaticity. On the other hand, in the experiments using Ne and methanol-ethanol the Raman bands remain narrow up to 36 GPa and 21 GPa (the maximum pressure reached), respectively. Therefore, uniaxial stresses are not noticeable in these experiments. This is consistent with the fact that no phase transition is detected up to 21 GPa in the methanol-ethanol experiment. It is known that the methanol-ethanol mixture provides better quasi-hydrostatic conditions compared to paraffin [22, 35]. Similar conclusions have been drawn recently from x-ray diffraction studies in the related material BaWO$_4$ [36]. Therefore, all the above described facts suggest that non-hydrostaticity could play a key role on the acceleration of the phase transition in wolframites. These results explain why in previous x-ray diffraction experiments performed on MgWO$_4$ in silicone oil [6], the onset of the transition from wolframite to the HP phase occurs around 11 GPa lower compared to the Ne experiments of this study; i.e., the results obtained using silicone oil are similar to those obtained using paraffin because of the lack of good hydrostaticity of both PTM above 10 GPa. It is commonly accepted the use ruby fluorescence to check hydrostaticity in DAC experiments [22]. Therefore, to further check the non-hydrostaticity hypothesis, we have also followed the FWHM of the fluorescent $R_1$ line of ruby in the experiments performed in Ne as well as without PTM. The obtained results support the conclusions about non-hydrostaticity derived from the analysis of Raman modes in $MgWO_4$. The FWHM of the $R_1$ line is the same in both experiments up to 5 GPa (0.5 nm). Beyond this pressure it increases in both experiments but at a very different rate. In the experiment with no PTM it grows up to 2 nm at 31 GPa while for the Ne experiment it is constrained to be smaller than 0.9 nm. This fact confirms that different stress distributions are present within the pressure chamber in different experiments.

To conclude this work, we would like to comment on the structure of the HP phase of $MgWO_4$. Previous calculations and Raman experiments on $ZnWO_4$ [8] and x-ray diffraction studies on $ZnWO_4$ and $MgWO_4$ [6] proposed the following HP structural sequence: $P2/c \rightarrow P\bar{1} \rightarrow C2/c \rightarrow Cmca$. However, the structure of the HP phase of wolframites is not fully determined yet. According to calculations, wolframite and a triclinic structure ($P\bar{1}$) are energetically competitive from 1 atm to 30 GPa [6]. A transition between both phases could be triggered by uniaxial stresses and would not involve any volume change. In addition, the x-ray powder diffraction pattern of the HP phase of $MgWO_4$ cannot be indexed with the monoclinic $C2/c$ structure and can be well explained considering the $P\bar{1}$ one. For a better identification of the HP phase of $MgWO_4$, the experimental results are compared with the calculated Raman modes of the two HP phase candidates ($P\bar{1}$ and $C2/c$) in Table IV. According to Table IV, it seems reasonable to affirm that the high pressure phase of $MgWO_4$ better resembles the triclinic structure than the monoclinic ($C2/c$) one. In particular, the calculated low- and high-frequency modes match well with the experimentally measured ones within 5%.

The same accounts for the pressure coefficients. However, the agreement is not so good for the modes of intermediate frequencies. In Fig. 9, we compare Raman spectra of the HP phases of Mg, Zn, and Cd wolframites with that of the triclinic phase of CuWO$_4$ at ambient pressure. We have also added ticks corresponding to the calculated modes for $C2/c$ and $P\bar{1}$ structures of MgWO$_4$. The Raman spectrum of HP-MgWO$_4$ shows more similarities with that of triclinic CuWO$_4$ than with those of monoclinic ($C2/c$) HP-CdWO$_4$ and HP-ZnWO$_4$. The only difference between the Raman spectra of HP-MgWO$_4$ and CuWO$_4$ is the shifting of Raman modes towards higher frequencies due to compression in HP-MgWO$_4$. In both cases there is a group of twelve modes at low frequencies plus three pair of modes at high frequencies associated to internal vibrations of the WO$_6$ octahedra. These observations provide additional support to the hypothesis that the HP phase of MgWO$_4$ could have a triclinic structure. However, further structural studies are requested to fully confirm it.

**V. SUMMARY**

In summary, we have performed Raman experiments on MgWO$_4$ using four different pressure-transmitting media with different hydrostaticity (Ne, methanol-ethanol, paraffin, and no medium at all). We detected a phase transition and determined the pressure dependence of the Raman modes of the low- and high-pressure phases. We also observed that non-hydrostatic conditions strongly affect the phase transition onset and the range of pressures at which coexistence of HP and LP phases occurs. Moreover, we have performed calculations that support our experimental conclusions and have helped us with the mode assignment and the identification of the possible structure of the high-pressure phase. We tentatively propose that the HP phase of MgWO$_4$ has a triclinic structure similar to that of CuWO$_4$. In addition, we have reported Raman measurements in MnWO$_4$, FeWO$_4$, CoWO$_4$, and NiWO$_4$ at ambient pressure and provided calculations for these compounds both at ambient and high pressures. A systematic comparison between theory and experiments is presented for the whole family of wolframites and the effect of the divalent cation on the Raman frequencies is discussed.

**Acknowledgements**

Research was financed by the Spanish MEC under Grants No. MAT2010-21270-C04-01/02/04, and No. CSD-2007-00045. J.R.-F. thanks the support from the MEC through the FPI program as well as the SPP1236 central facility in Frankfurt for its use. F.J.M. acknowledges support from Vicerrectorado de Investigación y Desarrollo de la UPV (Grant No. UPV2010-0096). A.M. and P.R.-H. acknowledge the supercomputer time provided by the Red Española de Supercomputación (RES). A.F. appreciates support from the German Research Foundation (Grant No. FR2491/2-1).

**Table I.** Calculated and experimental [6] Raman modes together with their pressure coefficients and Grüneisen parameters for the wolframite $P2/c$ phase of MgWO$_4$.

| Mode | Ab initio | | | Experiment | | |
|---|---|---|---|---|---|---|
| | $\omega$ (cm$^{-1}$) | $d\omega/dP$ (cm$^{-1}$GPa$^{-1}$) | $\gamma$ | $\omega$ (cm$^{-1}$) | $d\omega/dP$ (cm$^{-1}$GPa$^{-1}$) | $\gamma$ |
| B$_g$ | 104.3 | 0.80 | 1.23 | 97.4 | 0.69 | 0.93 |
| A$_g$ | 152.1 | 0.24 | 0.25 | 155.9 | 0.26 | 0.22 |
| B$_g$ | 184.6 | 0.44 | 0.38 | 185.1 | 0.51 | 0.36 |
| B$_g$ | 215.3 | 0.62 | 0.46 | 215.0 | 0.63 | 0.38 |
| B$_g$ | 267.7 | 1.01 | 0.61 | 266.7 | 1.01 | 0.50 |
| A$_g$ | 287.0 | 0.51 | 0.29 | 277.1 | 0.55 | 0.26 |
| A$_g$ | 301.5 | 1.93 | 1.03 | 294.1 | 1.92 | 0.86 |
| B$_g$ | 308.8 | 1.79 | 0.93 | 313.9 | 1.99 | 0.83 |
| A$_g$ | 361.8 | 4.20 | 1.87 | 351.9 | 3.52 | 1.31 |
| B$_g$ | 372.3 | 3.90 | 1.69 | 384.8 | 4.95 | 1.69 |
| B$_g$ | 405.2 | 5.42 | 2.15 | 405.2 | 1.47 | 0.48 |
| A$_g$ | 411.3 | 1.67 | 0.65 | 420.4 | 1.59 | 0.50 |
| B$_g$ | 523.4 | 3.31 | 1.02 | 518.1 | 3.30 | 0.84 |
| A$_g$ | 560.9 | 3.33 | 0.96 | 551.6 | 3.00 | 0.71 |
| B$_g$ | 683.2 | 4.34 | 1.02 | 683.9 | 4.09 | 0.78 |
| A$_g$ | 720.7 | 3.34 | 0.75 | 713.2 | 3.35 | 0.62 |
| B$_g$ | 809.8 | 4.14 | 0.82 | 808.5 | 3.69 | 0.60 |
| A$_g$ | 912.5 | 3.61 | 0.64 | 916.8 | 3.19 | 0.46 |

**Table II.** Experimental and calculated Raman modes in cm$^{-1}$ of all wolframite compounds. The experimental data of MgWO$_4$, FeWO$_4$, CoWO$_4$, and NiWO$_4$ are from the present study. The experimental pressure coefficients [$d\omega/dP$ (cm$^{-1}$/GPa)] are also included in parenthesis for those compounds that are available. The internal modes are denoted by an asterisk.

| Mode | MgWO$_4$ [6] Exp. | MgWO$_4$ [6] Calc. | Mode | MnWO$_4$ [22, 23, 24] Exp. | MnWO$_4$ [22, 23, 24] Calc. | ZnWO$_4$ [8] Exp. | ZnWO$_4$ [8] Calc. | CdWO$_4$ [9] Exp. | CdWO$_4$ [9] Calc. |
|---|---|---|---|---|---|---|---|---|---|
| B$_g$ | 97.4 (0.69) | 104 (0.80) | B$_g$ | 89 (0.81) | 95 (0.78) | 91.5 (0.95) | 84 (1.02) | 78 (0.52) | 67 (0.89) |
| A$_g$ | 155.9 (0.26) | 152 (0.24) | A$_g$ | 129 (0.25) | 129 (-0.06) | 123.1 (0.65) | 119 (0.48) | 100 (0.69) | 97 (0.36) |
| B$_g$ | 185.1 (0.51) | 185 (0.44) | B$_g$ | 160 | 165 (0.27) | 145.8 (1.20) | 137 (1.33) | 118 (1.02) | 111 (0.91) |
| B$_g$ | 215.0 (0.63) | 215 (0.62) | B$_g$ | 166 (0.96) | 171 (0.54) | 164.1 (0.72) | 163 (0.42) | 134 (0.82) | 126 (0.74) |
| B$_g$ | 266.7 (1.01) | 268 (1.01) | B$_g$ | 177 | 183 (0.72) | 189.6 (0.67) | 182 (0.41) | 148 (1.51) | 142 (1.03) |
| A$_g$ | 277.1 (0.55) | 287 (0.51) | A$_g$ | 206 (2.36) | 226 (2.19) | 196.1 (2.25) | 186 (2.52) | 177 (0.71) | 177 (0.70) |
| A$_g$ | 294.1 (1.92) | 302 (1.93) | B$_g$ | 272 | 278 (1.82) | 267.1 (1.32) | 261 (2.16) | 249 (2.14) | 239 (1.86) |
| B$_g$ | 313.9 (1.99) | 309 (1.79) | A$_g$ | 258 (0.22) | 264 (0.34) | 276.1 (0.87) | 264 (0.82) | 229 (0.29) | 220 (0.11) |
| A$_g$ | 351.9 (3.52) | 362 (4.20) | B$_g$ | 294 (1.74) | 296 (2.72) | 313.1 (1.74) | 298 (1.44) | 269 (1.41) | 252 (1.70) |
| B$_g$' | 384.8 (4.95) | 372 (4.90) | A$_g$ | 327 | 338 (2.4) | 342.1 (1.74) | 324 (1.70) | 306 (0.04) | 287 (0.12) |
| B$_g$ | 405.2 (1.47) | 405 (5.42) | B$_g$' | 356 | 373 (4.6) | 354.1 (3.87) | 342 (3.3) | 352 (4.55) | 338 (4.14) |
| A$_g$* | 420.4 (1.59) | 411 (1.67) | A$_g$* | 397 (1.64) | 389 (1.71) | 407 (1.65) | 384 (1.84) | 388 (2.33) | 357 (2.39) |
| B$_g$ | 518.1 (3.30) | 523 (3.31) | B$_g$ | 512 (2.99) | 509 (2.93) | 514.5 (3.18) | 481 (3.1) | 514 (3.86) | 490 (2.63) |
| A$_g$* | 551.6 (3.00) | 561 (3.33) | A$_g$* | 545 (2.79) | 548 (2.77) | 545.5 (3.00) | 515 (3.07) | 546 (2.32) | 531 (1.51) |
| B$_g$* | 683.9 (4.09) | 683 (4.34) | B$_g$* | 674 | 662 (3.79) | 677.8 (3.90) | 636 (3.90) | 688 (4.35) | 656 (3.68) |
| A$_g$* | 713.2 (3.35) | 721 (3.34) | A$_g$* | 698 (3.25) | 694 (2.75) | 708.9 (3.30) | 679 (3.24) | 707 (3.92) | 684 (3.32) |
| B$_g$* | 808.5 (3.69) | 810 (4.14) | B$_g$* | 774 (3.83) | 775 (3.54) | 786.1 (4.40) | 753 (4.00) | 771 (4.30) | 743 (3.95) |
| A$_g$* | 916.8 (3.19) | 913 (3.61) | A$_g$* | 885 (2.20) | 858 (1.82) | 906.9 (3.70) | 862 (3.36) | 897 (3.66) | 864 (2.88) |

**Table II.** Continuation

| Mode | FeWO$_4$ | | CoWO$_4$ | | NiWO$_4$ | | CuWO$_4$ [10] | |
|---|---|---|---|---|---|---|---|---|
| | Exp. | Calc. | Exp. | Calc. | Exp. | Calc. | Exp. | Calc. |
| B$_g$ | 86 | 92 (1.13) | 88 | 99 (0.63) | 91 | 103 (0.63) | 97 (0.90) | 90 (0.75) |
| A$_g$ | 124 | 132 (0.22) | 125 | 140 (0.17) | 141 | 147 (0.11) | 129 (-0.09) | 111 (0.32) |
| B$_g$ | 154 | 162 (1.50) | 154 | 168 (0.52) | 165 | 180 (0.55) | 157 (0.87) | 154 (0.19) |
| B$_g$ | 174 | 179 (0.59) | 182 | 192 (0.21) | 190 | 207 (0.56) | 178 (0.45) | 173 (1.01) |
| B$_g$ | | 184 (1.06) | 199 | 193 (0.59) | 201 | 220 (0.79) | 190 (0.43) | 186 (0.39) |
| A$_g$ | 208 | 213 (3.81) | | 237 (2.37) | | 246 (2.24) | 192 (2.50) | 203 (0.38) |
| B$_g$ | 266 | 263 (3.17) | 271 | 292 (2.27) | 298 | 305 (2.73) | 275 (1.34) | 215 (0.44) |
| A$_g$ | | 278 (0.67) | | 299 (0.52) | 298 | 313 (0.75) | 285 (2.45) | 265 (0.67) |
| B$_g$ | 299 | 295 (1.55) | 315 | 317 (1.96) | | 329 (1.68) | 312 (1.48) | 285 (0.09) |
| A$_g$ | 330 | 330 (2.73) | 332 | 361 (3.46) | 354 | 382 (3.18) | 316 (1.58) | 315 (0.29) |
| B$_g$' | | 350 (4.27) | | 379 (3.99) | | 402 (3.85) | 367 (3.57) | 331 (0.31) |
| A$_g$* | 401 | 406 (0.54) | 403 | 407 (1.15) | 412 | 416 (1.31) | 391 (1.33) | 459 (1.43) |
| B$_g$ | 500 | 483 (3.47) | 496 | 510 (2.9) | 505 | 512 (2.99) | 505 (3.30) | 502 (3.57) |
| A$_g$* | 534 | 530 (3.48) | 530 | 551 (3.25) | 537 | 557 (3.33) | 548 (2.97) | 561 (2.57) |
| B$_g$* | 653 | 637 (4.77) | 657 | 646 (3.81) | 663 | 671 (3.85) | 645 (3.74) | 699 (2.78) |
| A$_g$* | 692 | 676 (3.75) | 686 | 692 (3.01) | 688 | 702 (3.07) | 686 (2.99) | 745 (1.77) |
| B$_g$* | 777 | 754 (4.50) | 765 | 781 (3.88) | 765 | 786 (4.08) | 749 (4.04) | 919 (2.12) |
| A$_g$* | 878 | 866 (3.68) | 881 | 874 (2.80) | 887 | 881 (3.17) | 847 (3.09) | 963 (2.53) |

**Table III.** Experimental and calculated crystal parameters for wolframite $P2/c$ phase of FeWO$_4$, CoWO$_4$, and NiWO$_4$.

| Cell parameters and Wychoff positions | FeWO$_4$ Exp. [17] | FeWO$_4$ Calc. | CoWO$_4$ Exp. [18] | CoWO$_4$ Calc. | NiWO$_4$ Exp. [18] | NiWO$_4$ Calc. |
|---|---|---|---|---|---|---|
| a (Å) | 4.753 | 4.7889 | 4.6698 | 4.5583 | 4.5992 | 4.5104 |
| b (Å) | 5.720 | 5.8278 | 5.6873 | 5.6183 | 5.6606 | 5.5842 |
| c (Å) | 4.968 | 5.0165 | 4.9515 | 4.8908 | 4.9068 | 4.8608 |
| β (deg.) | 90.08 | 90.43 | 90.00 | 89.580 | 90.03 | 89.625 |
| A cation site: 2f (1/2, y, 1/4) | 0.6784 | 0.6698 | 0.6712 | 0.6587 | 0.6616 | 0.6542 |
| W site: 2e (0, y, 1/4) | 0.1808 | 0.1795 | 0.1773 | 0.1792 | 0.1786 | 0.1792 |
| O$_1$ site: 4g (x, y, z) | 0.2167 | 0.2134 | 0.2176 | 0.2217 | 0.2241 | 0.2245 |
|  | 0.1017 | 0.1043 | 0.1080 | 0.1086 | 0.1105 | 0.1096 |
|  | 0.5833 | 0.5650 | 0.9321 | 0.9269 | 0.9204 | 0.9263 |
| O$_2$ dite: 4g (x, y, z) | 0.2583 | 0.2532 | 0.2540 | 0.26212 | 0.2644 | 0.2623 |
|  | 0.3900 | 0.3742 | 0.3757 | 0.37847 | 0.3772 | 0.3802 |
|  | 0.0900 | 0.1076 | 0.3939 | 0.40338 | 0.3953 | 0.4076 |

**Table IV.** Experimental and calculated Raman frequencies considering $P\bar{1}$ and $C2/c$ structures as well as pressure coefficients for the high-pressure phase of MgWO$_4$.

| | *Ab initio* $P\bar{1}$ at 30.5 GPa | | | *Ab initio* $C2/c$ at 38.4 GPa | | Experiment at 30.6 GPa | |
|---|---|---|---|---|---|---|---|
| Mode | ω (cm$^{-1}$) | dω/dP (cm$^{-1}$GPa$^{-1}$) | Mode | ω (cm$^{-1}$) | dω/dP (cm$^{-1}$GPa$^{-1}$) | ω (cm$^{-1}$) | dω/dP (cm$^{-1}$GPa$^{-1}$) |
| A$_g$ | 124.3 | 0.40 | B$_g$ | 177.9 | 0.24 | 166 | 0.60 |
| A$_g$ | 155.6 | 0.00 | A$_g$ | 217.5 | 0.00 | 185 | 2.01 |
| A$_g$ | 193.0 | 0.01 | B$_g$ | 220.2 | 0.38 | 210 | 1.52 |
| A$_g$ | 235.5 | 0.82 | A$_g$ | 271.0 | 0.71 | 239 | 0.71 |
| A$_g$ | 291.9 | 0.93 | B$_g$ | 295.5 | 0.71 | 276 | 2.18 |
| A$_g$ | 300.7 | 0.47 | A$_g$ | 377.9 | 1.59 | 345 | 1.62 |
| A$_g$ | 343.2 | 0.59 | B$_g$ | 391.0 | 1.14 | 383 | 1.89 |
| A$_g$ | 367.3 | 2.45 | A$_g$ | 394.7 | 3.39 | 393 | 1.09 |
| A$_g$ | 457.0 | 1.62 | B$_g$ | 423.4 | 2.88 | 420 | 2.02 |
| A$_g$ | 474.6 | 2.52 | A$_g$ | 490.2 | 2.85 | 441 | 2.46 |
| A$_g$ | 475.8 | 2.55 | B$_g$ | 528.3 | 2.2 | 531 | 2.75 |
| A$_g$ | 551.7 | 3.56 | B$_g$ | 609.9 | 2.53 | 542 | 0.88 |
| A$_g$ | 619.5 | 3.34 | B$_g$ | 647.2 | 2.60 | 609 | 0.62 |
| A$_g$ | 650.0 | 2.53 | A$_g$ | 717.3 | 2.18 | 700 | 2.91 |
| A$_g$ | 805.4 | 3.46 | A$_g$ | 834.9 | 1.95 | 724 | 2.63 |
| A$_g$ | 815.8 | 2.84 | B$_g$ | 877.6 | 2.21 | 801 | 2.63 |
| A$_g$ | 939.6 | 2.96 | A$_g$ | 918.7 | 2.70 | 833 | 3.26 |
| A$_g$ | 1016.6 | 2.23 | B$_g$ | 942.1 | 3.03 | 971 | 1.45 |

**Figure Captions**

**Figure 1.** (color online) MgWO$_4$ wolframite. Big spheres are W atoms, middle-sized ones are Mg atoms, and small ones are O atoms.

**Figure 2.** Raman spectra of wolframite MgWO$_4$ at selected pressures (Ne experiment). Arrows indicate the appearance of the strongest band of the HP phase and the strongest band of the wolframite phase after the transition onset. All spectra are measured upon pressure increase with the exception of those denoted by (r), which correspond to pressure release.

**Figure 3.** Raman spectra of wolframite MgWO$_4$ at different pressures. (a) No PTM, (b) spectroscopic paraffin. Arrows indicate the appearance of the strongest high-pressure peak and the strongest peak of the low-pressure phase after the transition onset.

**Figure 4.** Pressure dependence of the Raman mode frequencies of the wolframite (solid symbols) and high-pressure (empty symbols) phases of MgWO$_4$ and linear fittings. Circles denote methanol-ethanol experiments while triangles belong to Ne experiments. The vertical dashed line indicates the onset of the phase transition.

**Figure 5.** Divalent-cation reduced-mass dependence of the ambient conditions Raman frequencies for MgWO$_4$, ZnWO$_4$ and CdWO$_4$. Stars indicate B$_g$ modes whereas the triangles refer to A$_g$ modes. Mode symmetries are indicated and the mode denoted as B$_g$´ in the text highlighted using bold characters.

**Figure 6.** Divalent-cation reduced-mass dependence of the ambient conditions Raman frequencies for MnWO$_4$, FeWO$_4$, CoWO$_4$ and NiWO$_4$. Stars indicate B$_g$ modes whereas the triangles refer to A$_g$ modes. Mode symmetries are indicated and the mode denoted as B$_g$´ in the text highlighted using bold characters.

**Figure 7.** Ratio between the intensity of the strongest peaks of wolframite ($I_{LP}$) and high-pressure ($I_{HP}$) phase of MgWO$_4$. Solid symbols indicate Ne experiments whereas the empty ones are from the paraffin experiment.

**Figure 8.** Full-width at half maximum of the highest (a) and lowest (b) frequency peaks of wolframite MgWO$_4$ at different pressures. Different symbols indicate different experimental conditions.

**Figure 9.** Raman spectra of the high-pressure phases of MgWO$_4$, ZnWO$_4$ [8] and CdWO$_4$ [9] wolframites at around 40 GPa. The tips indicate the *ab initio* calculated modes considering the triclinic ($P\bar{1}$) and monoclinic β-fergusonite (*C*2/*c*) phases at 30.5 and 38.4 GPa, respectively. The Raman spectrum of triclinic CuWO$_4$ at ambient pressure [10] is also included for comparison.

Figure 1

Figure 2

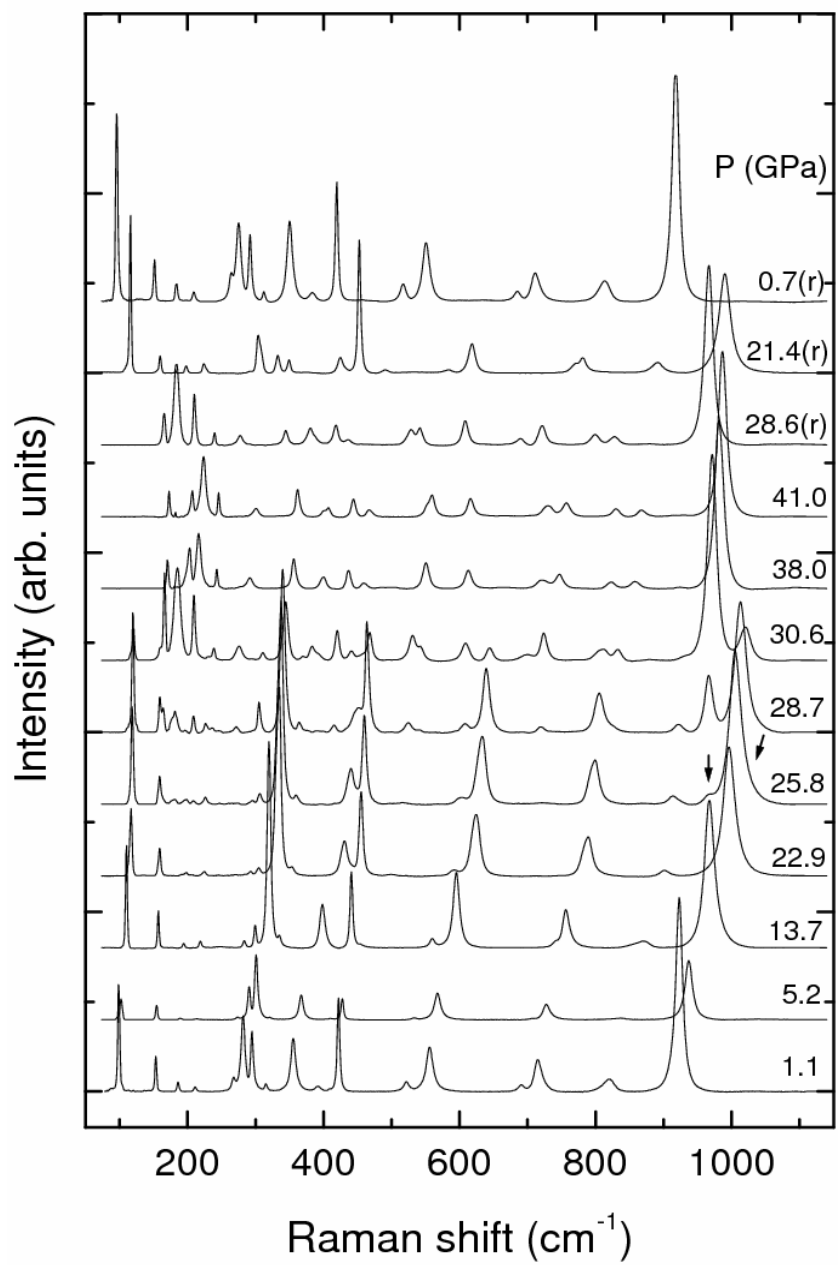

Figure 3

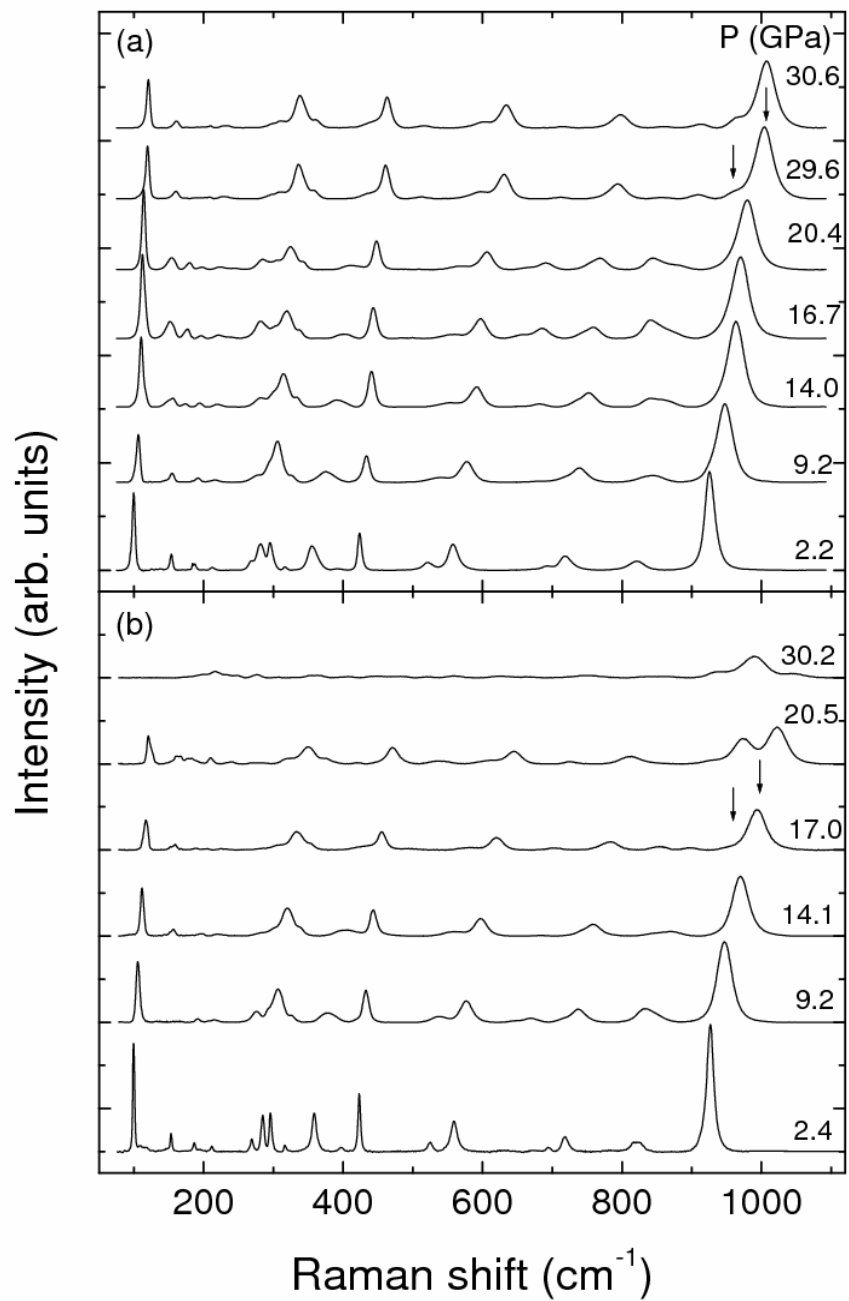

Figure 4

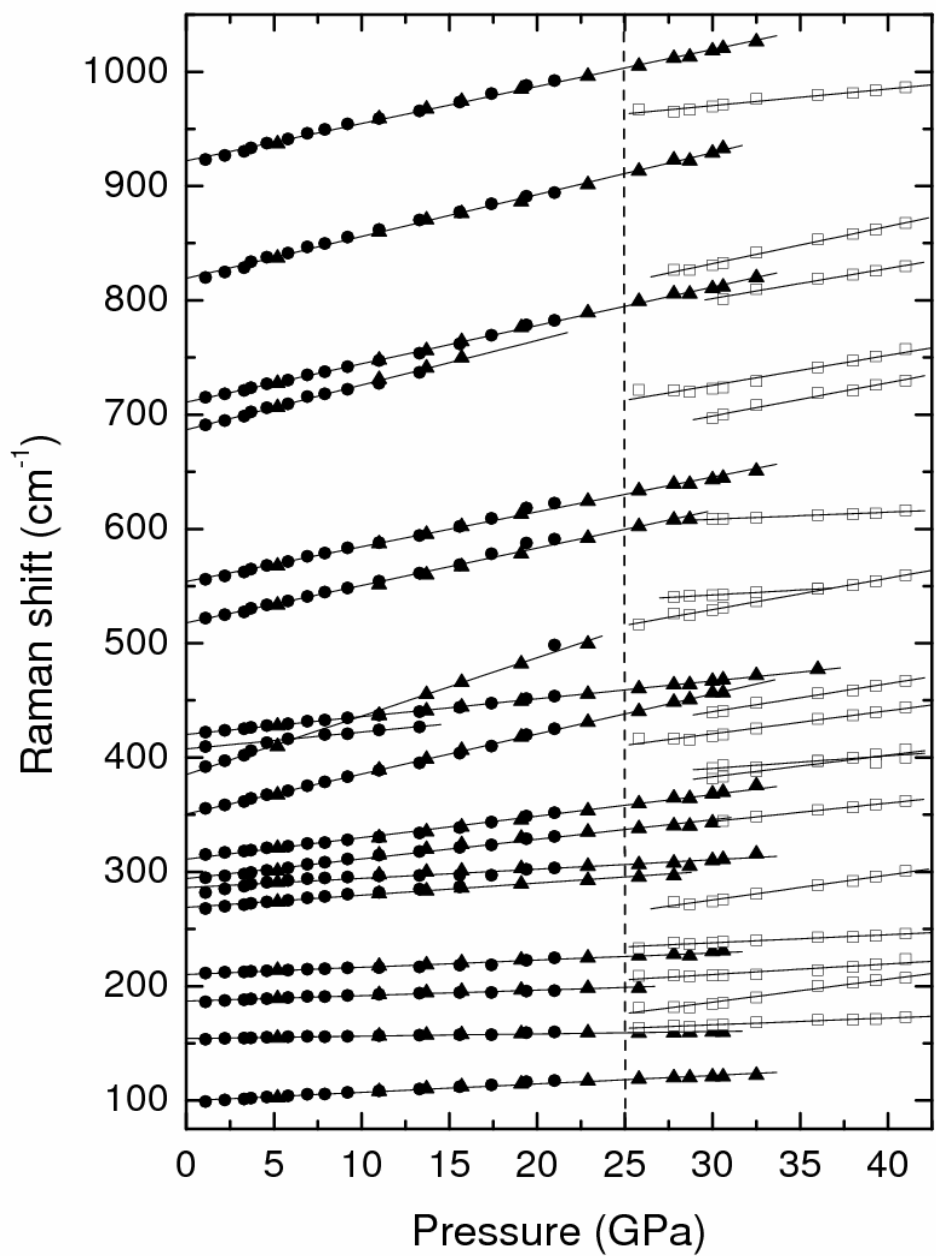

Figure 5

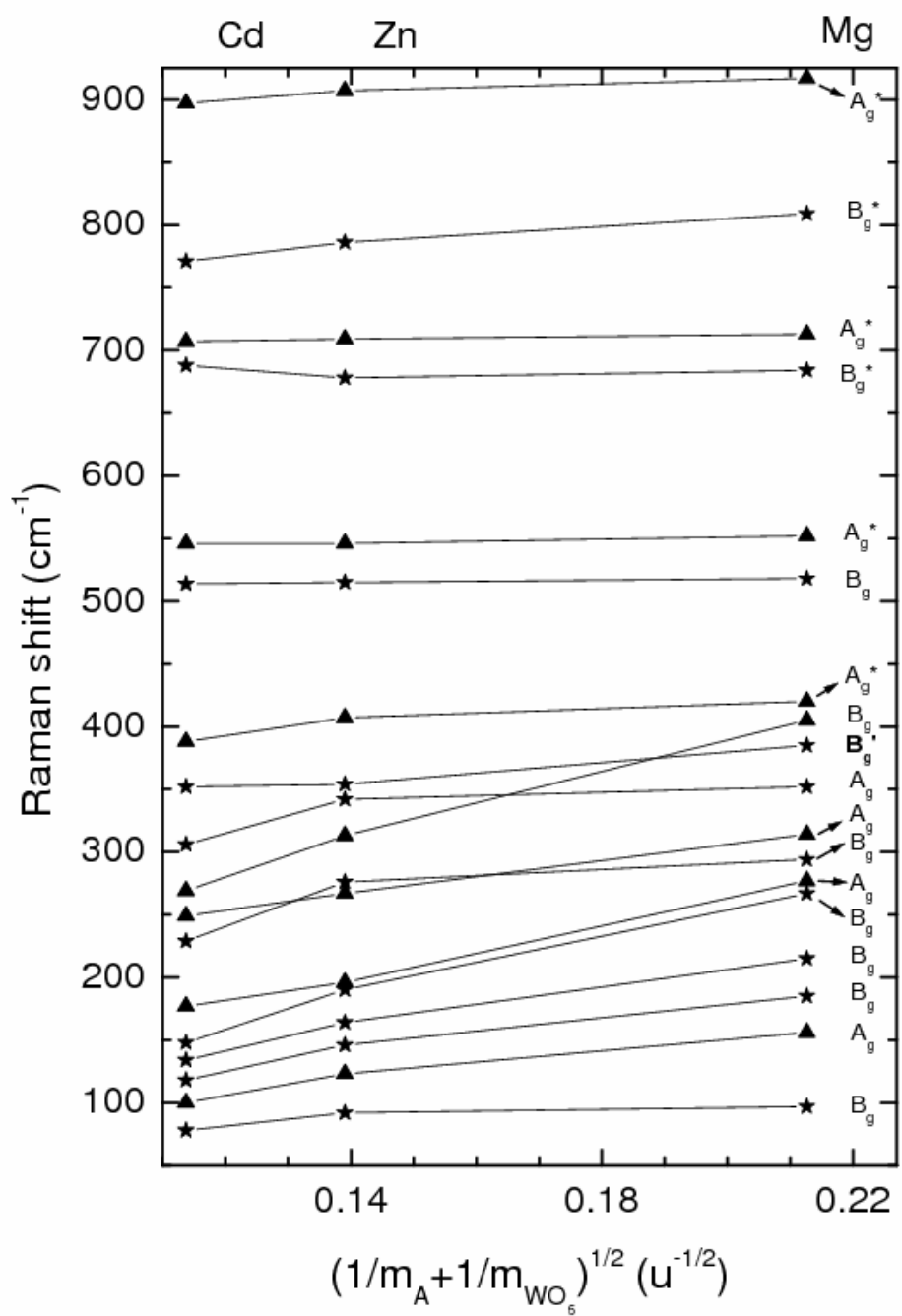

Figure 6

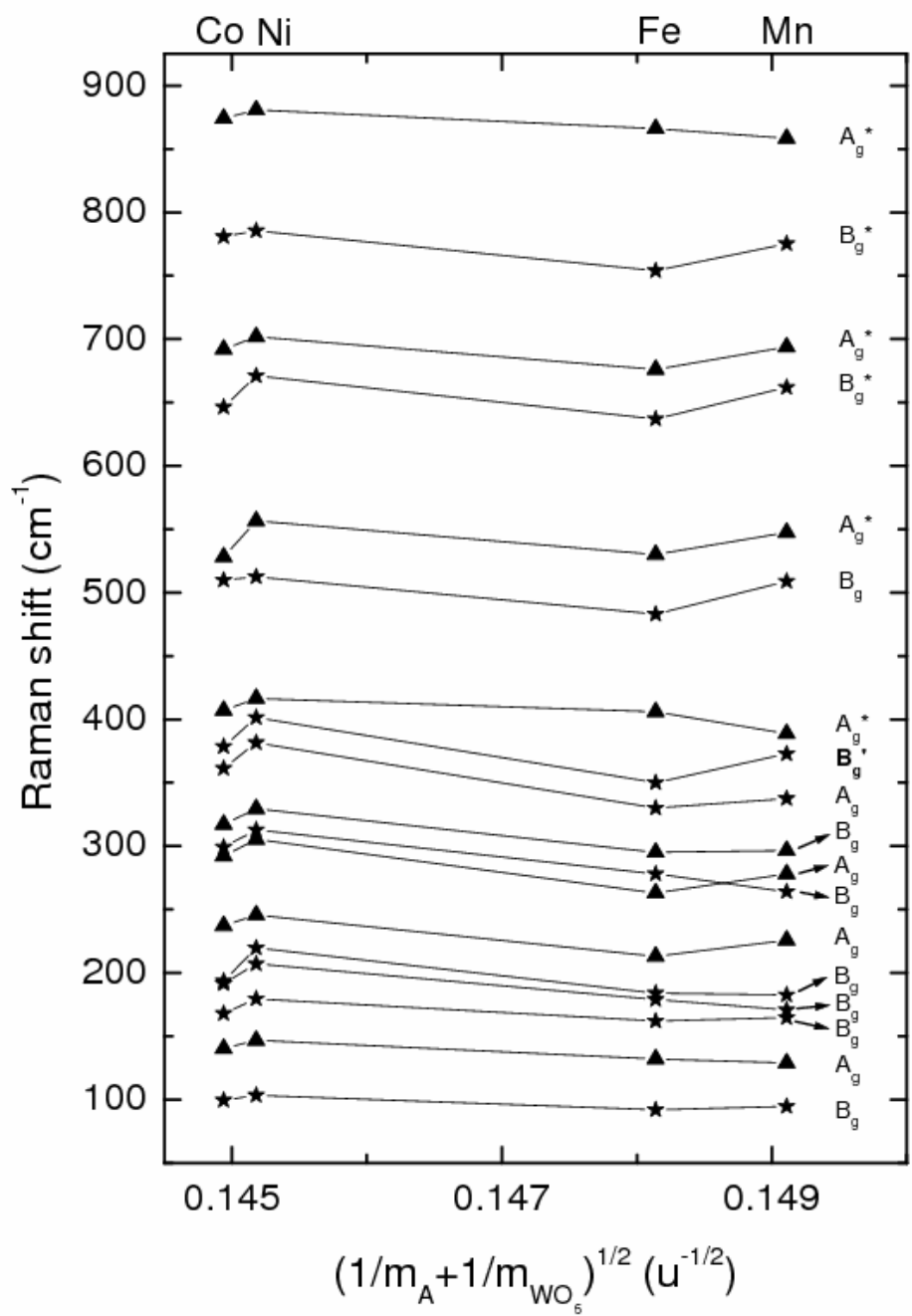

Figure 7

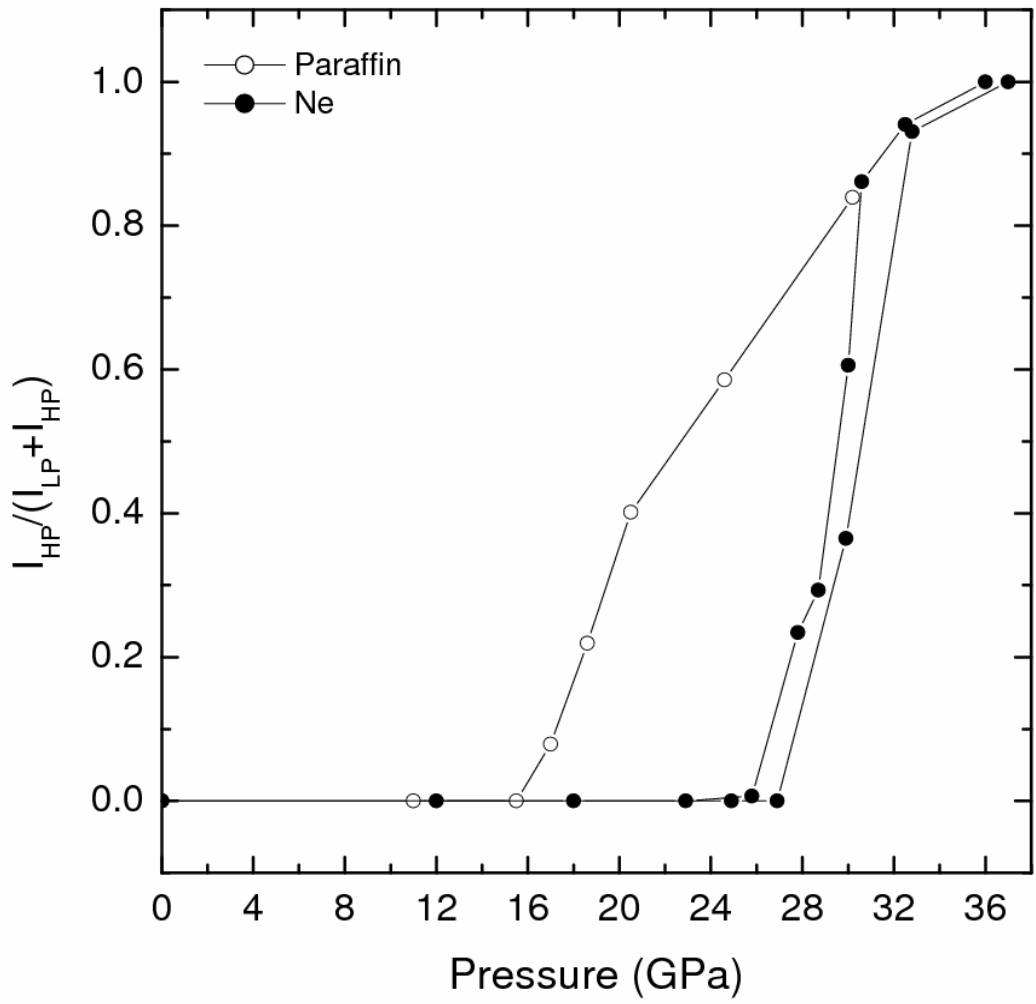

Figure 8

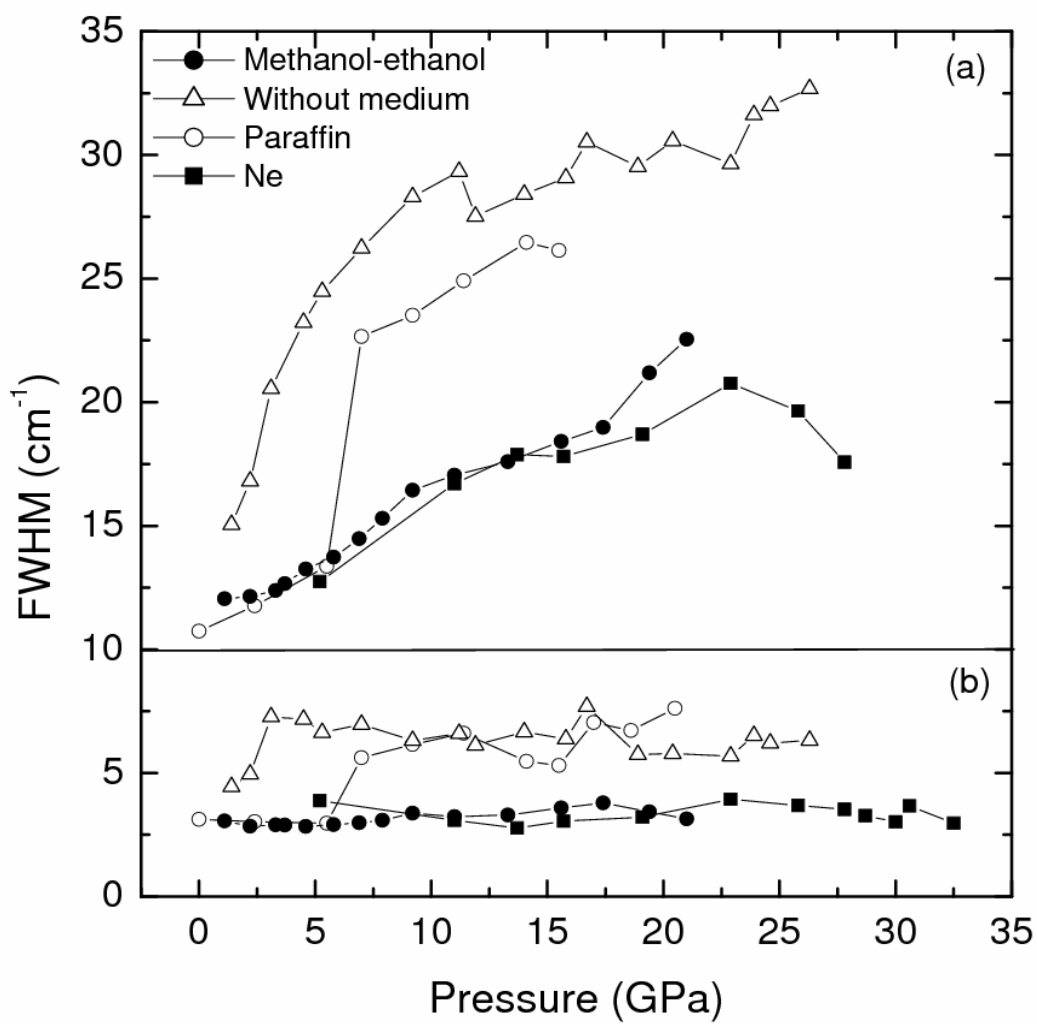

Figure 9

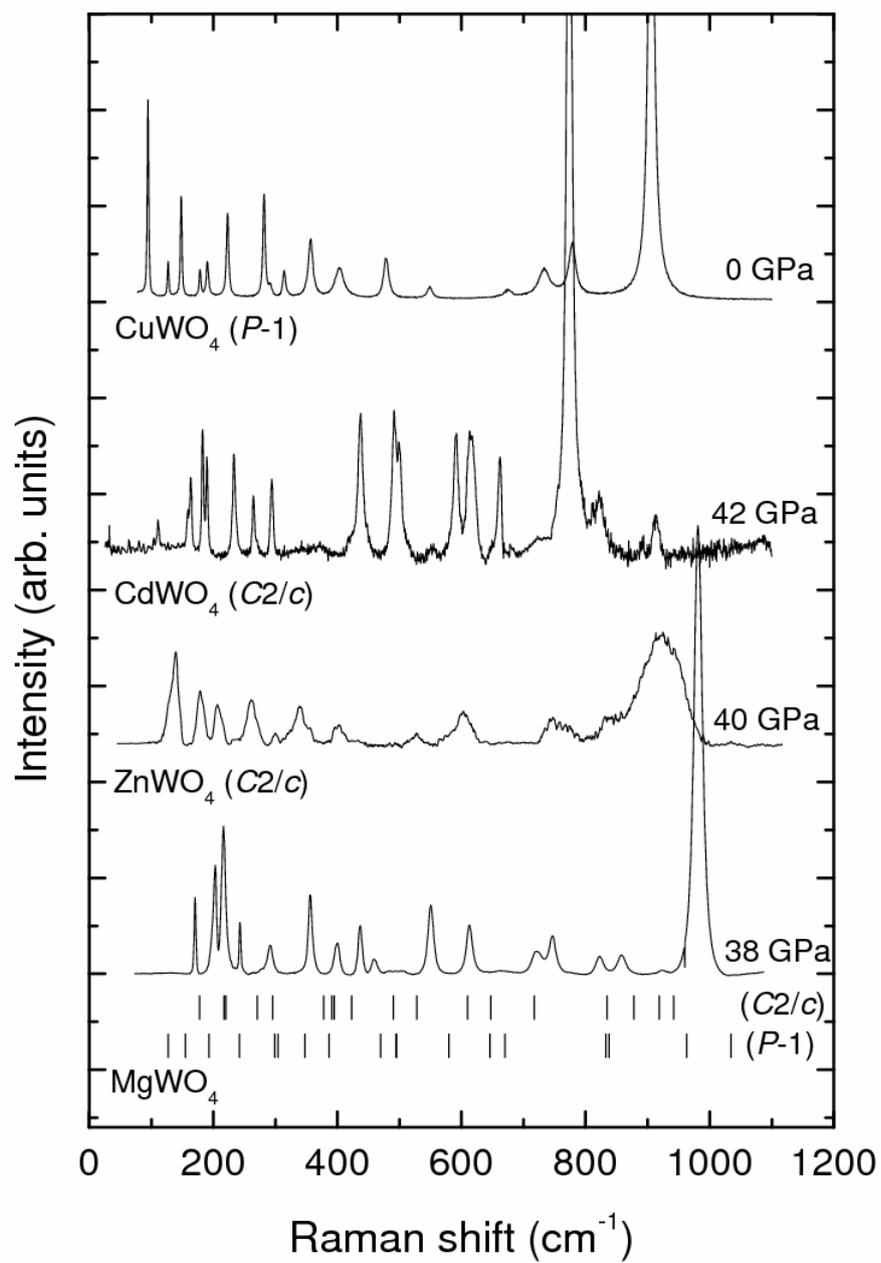